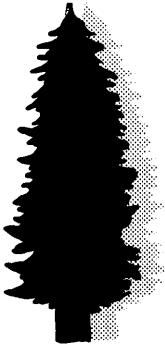

# IMPROVED WAYS TO COMPARE SIMULATIONS TO DATA*

Joel R. Primack
Santa Cruz Institute for Particle Physics,
University of California, Santa Cruz, CA 95064

**Abstract**

Theoretical models for structure formation with Gaussian initial fluctuations have been worked out in considerable detail and compared with observations on various scales. It is on nonlinear scales $\lesssim 10\ h^{-1}$ Mpc that the greatest differences exist between $\Omega = 1$ models that have been normalized to agree on the largest scales with the COBE data; here especially there is a need for better statistical tests which are simultaneously *robust*, *discriminatory*, and *interpretable*. The era at which galaxy and cluster formation occurs is also a critical test of some models. Needs for the future include faster and cleverer codes, better control of cosmic variance in simulations, better understanding of processes leading to galaxy formation, better ways of comparing observational data with models, and better access to observational and simulation data.

## 1 Introduction

Although many cosmological models have been considered by various authors, I propose to concentrate here on a particular class of such models, namely those inspired by the hypotheses of inflation (hence with $\Omega = 1$, or at least curvature $k = 0$, and a near-Zel'dovich primordial spectrum of adiabatic fluctuations) and (all or mostly) cold dark matter. I do this not only because I believe that such models still have the best prospects of ultimately being found to be consistent with the data. My main motivation for concentrating on these models is that they are *well specified*, in the sense that they are described by a small number of adjustable parameters (unlike general non-Gaussian models, for example), and very *predictive*, in the sense that many of their consequences can be worked out fairly easily with relatively few uncontrolled approximations. Thus they can be confronted rather directly with the observational data, and eventually most (or all) such theories can actually be ruled out — as standard CDM already

---

*To be published in the *Proceedings of the XXX Rencontres de Moriond: Clustering in the Universe*, edited by S. Maurogordato, C. Balkowski, C. Tao and Tran Thanh Van (Editions Frontieres, 1995).

has been (or is close to being) ruled out. If a small class of such models survive, they may actually have something to do with the real universe. Even if not, this research should help to develop better statistical methods for comparison of cosmological theories with observational data.

The great advantage of keeping a tentative theory in mind as one thinks about data is that it helps in organizing the facts. If it is a good theory, it will also call attention to particularly important facts — especially those that may contradict it! CDM stimulated the creation of better models of the origin and evolution of galaxies and large scale structure, it helped motivate the acquisition and analysis of crucial data, and it has been a valuable test bed for data analysis tools — allowing, for example, development and testing of the POTENT algorithm [1] for reconstructing the total density field from measured peculiar velocities. Comparison of the original standard $\Omega = 1$ CDM model and its variants (cf. e.g. [2]) with the observational data has certainly been useful during the past decade.

## 2 Testing Models

It is useful to divide the discussion of how to confront models with the observational data according to the scale of the observations: (a) greater than 100 Mpc, (b) 10-100 Mpc, (c) less than 10 Mpc, and (d) early structure formation. If we restrict attention to CDM-like $\Omega = 1$ models, the data on the largest scales (a) is probably useful mainly for establishing the normalization of the fluctuations, measuring the contribution of gravity waves, and testing inflation. For low-$\Omega$ CDM models with a cosmological constant $\Lambda$ such that the curvature vanishes (i.e. with $\Omega_\Lambda \equiv \Lambda/(3H_0^2) = 1 - \Omega_0$), data on the power spectrum $P(k)$ of galaxies and especially of the mass will be a crucial test, since if normalized to COBE [3] these models predict much higher $P(k)$ than $\Omega = 1$ CDM; however, the available data on the largest scales (mainly from the APM angular correlations and from the CfA2 + SSRS2 data [4]) is not yet a powerful constraint on theories. Comparing with the data on smaller scales tests the gravitational clustering hypothesis, and (assuming gravitational clustering is valid) the spectral shape and other features of the cosmological model, including the cosmological parameters $H_0$, $\Lambda$, and $\Omega$, and the nature of the dark matter — e.g., whether it is a mixture of cold and hot dark matter. With a given normalization of the spectrum, the smaller scale data also tests the shape of the primordial spectrum (e.g., whether there is "tilt"). But there are many problems with actually carrying out this program even when large-scale redshift surveys become available, perhaps the worst of which is the uncertainties about galaxy formation which make identification of "galaxies" in the simulations uncertain. Fortunately, there are several ingenious new techniques that promise to improve this situation. For example, weak gravitational lensing [5] or extension of peculiar velocity surveys to larger scales (which requires new ways of measuring distance independent of redshift, or of measuring the peculiar velocity directly e.g. by the Sunyaev-Zel'dovich velocity term) may allow direct determination of the distribution of mass on large or even very large scales.

### 2.1 Very large scales

The shape of the power spectrum deduced from the two-year COBE data and other large-angle CBR measurements is consistent with a power-law primordial power spectrum $P(k) = Ak^{n_p}$ with $n_p \approx 1.1 \pm 0.3$, and with the rms amplitude quoted as the quadrupole for an $n_p = 1$ spectrum given by $Q_{\rm rms-ps} \approx 20\,\mu{\rm K}$ [3]. If a CDM spectrum is normalized to this amplitude for $h = 0.5$, it appears to be consistent with all the available data on large scale structure (LSS) down to scales of approximately 100 $h^{-1}$ Mpc (e.g., [6, 1, 7]). But the amplitude of the LSS is presently known at best to about $\pm 25\%$ — that is roughly the uncertainty on the large-scale bulk velocity, for example (e.g., [1]), which at least measures the mass power spectrum. The amplitude of the galaxy or cluster power spectrum has the further uncertainty that the ratio of the rms fluctuation amplitudes of these objects to that of the mass, i.e. the "bias" $b_{\rm gal} = (\frac{\delta\rho}{\rho})_{\rm gal}/(\frac{\delta\rho}{\rho})_{\rm mass}$, is known rather poorly. Indeed, the extent to which the bias of a given category of astronomical objects can be regarded as a constant even on a given scale is not very well tested.

Within these fairly large uncertainties, the consistency between the CMB data and the LSS data supports the validity of the gravitational clustering hypothesis. In order to really test this both this and the hypothesis of cosmic inflation, it will be necessary to do better. Perhaps the most important issue is the contribution of gravity waves. These tensor modes can contribute to the low spherical harmonics $\ell \lesssim 20$ of the CMB temperature fluctuations, but hardly at all to the higher $\ell$ ones; and they are of course irrelevant to the formation of structure, to which only the scalar mode density fluctuations contribute. In principle, the tensor contribution can be determined by comparing the large and small angle CMB anisotropies, but in practice this will require more accurate CMB measurements on small scales than are presently available, and also knowledge of $\Omega_b$ and the extent of ionization of the universe after the recombination era, both of which have a strong influence on the amplitude of the CMB fluctuations near the first Doppler peak, $\ell \sim 200$. For the time being, it is perhaps best to regard the COBE normalization $Q_{\rm rms-ps} \approx 20$ as an *upper* limit to the density fluctuation amplitude, since the tensor and scalar modes add in quadruature. One of the most urgent issues for CMB studies is to determine a *lower* limit on the density fluctuation amplitude by constraining the tensor mode amplitude. This will allow improved tests of the gravitational clustering hypothesis, and measurement of the "tilt" of the primordial fluctuation amplitude.

All the nearby surveys, such as the CfA2 survey with an effective depth of about 15,000 km/s, have found structures as large as the surveys themselves. This left open the possibility that still larger structures would be found by even larger surveys, which would contradict the gravitational clustering hypothesis (e.g. [8]). Although the very large scale periodicity of peaks in the galaxy distribution with a length scale of $\sim 135\, h^{-1}$ Mpc seen in the BEKS [9] pencil beam redshift survey was unexpected in any cosmological model [10], it is significant that no indications of still larger scales were seen in this data (preliminary reports indicate that pencil beams in different directions also have peaks with such separations but not such strong periodicity). Large scale redshift surveys are now in progress, notably the KOSS southern sky redshift survey and the ESO Key Project. Preliminary reports suggest that no larger structures have in fact been found (Kirshner characterizes this as "the end of greatness"), again supporting the gravitational clustering hypothesis. The much larger scale surveys just beginning — the Two Degree Survey at the AAT, and the Sloan Digital Sky Survey at the Apache Point Observatory — will be able to measure the sizes of these large structures and characterize their correlations, shapes, and other statistical properties. These will provide essential constraints on models of cosmic structure formation. These statistical properties appear on the basis of the data available at present to be consistent with the expectations from CDM-type models, but it remains to be seen whether this is true for topological defect models.

## 2.2 Large scale structure ($\sim 10 - 100$ Mpc)

On these scales a great deal of galaxy redshift data [11] and peculiar velocity data [1] is already available, although much of it remains unpublished. There are also several redshift surveys for optically selected clusters, and large-scale redshift surveys for X-ray selected clusters (which are likely to be less affected by projection effects and galactic obscuration) are now in progress. Comparison of the spectrum of fluctuations measured with this data and with the small-angle CMB data when it is available will eventually provide a test of the gravitational clustering hypothesis.

Comparison with specific theories must be done on the basis of nonlinear simulations since on these scales linear theory is no longer reliable. All the available tests — power spectra of galaxies and mass (a preliminary POTENT result), galaxy and cluster correlations, skewness and higher moments of the pdf — suggest that CDM, normalized to fit on scales of 100 Mpc and above, fits increasingly poorly on smaller scales: it has too much power. For example, CDM predicts far too many clusters [12], and it predicts that mass autocorrelations become negative for separations beyond about 30 $h^{-1}$ Mpc, while all measurements of cluster correlations show that they remain significantly positive for separations out to $\sim 50$ $h^{-1}$ Mpc. These correlations are sensitive to the slope of the power spectrum, and indicate a steeper decline with increasing $k$ than CDM.

## 2.3 Intermediate scales (less than 10 Mpc)

It is on scales less than about 10 Mpc that the greatest differences exist between $\Omega = 1$ CDM variants that have all been normalized to agree on the largest scales with the COBE data. It is these scales on which galaxy locations and velocities, as revealed by relatively dense redshift surveys (i.e., with fainter galaxies included), have the greatest potential to help discriminate between cosmological models, for example those containing more or less of various mixtures of cold and hot dark matter, with or without a cosmological constant. (Someday there may be enough galaxy peculiar velocities based on accurate distance measurements independent of redshift to allow these to be used to discriminate between theories on small scales, but for the time being it remains necessary to smooth peculiar velocity data over scales of at least 5 Mpc to overcome the large uncertainty in each such measurement.) The statistics that have been used for this purpose include N-point functions, the void probability function VPF and related functions, skewness and kurtosis coefficients $S_3$ and $S_4$, multifractal analyses, the genus density, etc. These statistics indicate that galaxies exhibit hierarchical scaling as expected in gravitational clustering models, but most of these statistics (with the possible exception of the VPF, see e.g. [13]) do not appear to be able to discriminate very efficiently in redshift space between alternative Gaussian models – although they may discriminate between these and non-Gaussian (e.g., [14]) al. 1993) or scale-dependent-bias models (e.g., [15]).

Simulations are of random patches of the universe, so comparisons with observational data must be statistical. There are broadly two different approaches to making such comparisons; one can work either in the "theoretical plane"– i.e., attempt to "correct" the data for selection effects, redshift space effects, etc. – or in the "observational plane" – i.e., "observe" the simulations. As computational power has grown, it has become increasingly advisable to observe the simulations rather than attempt to correct the data, since simulations have much more information – for example, the velocity as well as location of each object identified as a galaxy. Thus it is possible to construct magnitude-limited redshift surveys from simulations with no ambiguity in the conversion to redshift space, but it is more difficult to recover real space information from only redshift space data.

On the other hand, one should not underestimate the difficulties of simulating observational data. Perhaps the greatest problem is determining which objects in the simulations are to be identified with observed galaxies. In dissipationless simulations, with only dark matter included, perhaps the worst problem is overmerging. Nearby dark matter halos merge into large blobs, even though in the real universe the individual galaxies within a group or cluster can retain their separate identities since the gas can condense a great deal within the larger dark matter halos. Even in hydrodynamical simulations (e.g. [16]) there are serious limitations; only limited spatial resolution is available with even the largest supercomputers, and many relevant physical processes such as energy input from stars and supernovae are neglected or treated superficially. Although the main strengths and limitations of the several different approaches to hydrodynamical simulations seem to be reasonably well understood (see e.g. [17]), the accuracy and resolution currently available is limited.

The necessary art, at the present stage of cosmology, is to invent statistical tests that are both *robust* and *discriminatory*. Robust, in the sense that the difficulties of the sort mentioned above – e.g. in galaxy identification or "illumination" (assignment of luminosity to objects identified as galaxies) – do not significantly affect the statistical results. And discriminatory, in the sense that the statistical tests give significantly different results for the various cosmological models that are of interest. That a given statistical test is actually robust and discriminatory can be checked by trying a wide variety of different approaches to galaxy identification and illumination of simulations of a number of different cosmological models. A further desirable feature of cosmological statistics is *interpretability* in terms of the physical assumptions or observational consequences of the cosmological model in question. For example, the matter two-point correlation function is just the Fourier transform of the (nonlinear, i.e. evolved) power spectrum $P(k)$, which is of fundamental theoretical importance.

For examples of such statistics and tests by my collaborators and students, see e.g. [18, 19] (galaxy group statistics), [13] (void probability function), [20, 21] (shape statistics). To avoid the problem of cosmic variance, discussed in more detail in the next section, we should ideally have compared many cosmological models by running simulations of them with the same random numbers (producing for example the same random phases). In the test-bed of simulations that we had available this was only possible to a limited extent, but we are improving on this.

## 3   Large-Scale Constrained Realizations

A great deal of effort is being devoted to creating improved methods of doing dissipationless and hydrodynamical cosmological simulations. In a new research project with Avishai Dekel and our students and other collaborators, we propose to complement this by developing more efficient methods for setting up such simiulations, for comparing the results to observational data.

The distribution and velocities of galaxies on scales of $\sim 1 - 10 \ h^{-1}$ Mpc as revealed by redshift surveys are particularly sensitive to the nature of the dark matter, but discriminatory statistics such as relative galaxy velocities are also sensitive to the largest waves in the simulation volume. Because there are only a few such waves, they cannot fairly represent the Gaussian distribution assumed in models. Moreover, the dominant structures — rich clusters, "great walls" — in the largest dense redshift surveys such as PPS, CfA2, and SSRS2 also strongly affect statistics such as velocities. The solution we propose is to do simulations to be compared to **specific** redshift surveys, using the technique of constrained Gaussian realizations to set up initial conditions that will produce these dominant structures, with smaller waves put in and the simulation evolved to the present using the mixture of cosmological parameters ($H_0$, $\Omega$, $\Lambda$) and dark matter types according to each model to be tested. This "Large Scale Constrained Realizations" (LSCR) approach certainly needs much development and testing; but assuming that it works as well as we hope, we anticipate that it could grow into a major addition to the technology of observational cosmology.

Cosmic variance is perhaps the most serious problem in comparing simulations such as ours to redshift data. The cosmic scatter between random realizations is artificially large because the perturbations of the largest scales are represented by only a few waves and they therefore do not represent properly a Gaussian field. Each such realization is therefore typically dominated by one or a few large-scale waves, with strong systematic effects on the small-scale structure of interest, and especially on the velocities. A brute force way to beat this cosmic scatter is by averaging over many random realizations, but this can be quite expensive and impractical with full N-body simulations, although it is quite practical with the truncated Zel'dovich approximation (e.g. [22]).

A much more economical way would be to fix the large-scale structure in the initial conditions at its true pattern, and generate random realizations of the relevant small-scale structure only. The large waves on scales $> 20$ $h^{-1}$ Mpc in regions of our cosmological neighborhood covered by dense redshift surveys can be extracted from the IRAS redshift surveys or from peculiar velocity data (e.g. using the POTENT reconstruction [1]. These large-scale constraints will be imposed on a random Gaussian realization of smaller waves representing each of the models of interest using the technique of Hoffman [23].

We also need to test the effectiveness of the overall LSCR procedure. For example, we plan to impose LS constraints from both CHDM and $\Lambda$CDM simulations, use the LSCR procedure to set up initial conditions for each model with the same parameters, and then evolve all four models — CHDM-CHDM, $\Lambda$CDM-$\Lambda$CDM, and the two crossed models — to the present. We can then see how well we can recover the same statistical results on various tests as in the original CHDM and $\Lambda$CDM models, and understand the nature of the biases if any in the crossed models.

## 4 Early structure formation

A major difference between cosmological models is in their predictions for the origins of galaxies, clusters, and large scale structure, and the evolution of these with redshift. Detection of large-mass collapsed objects at high redshift would certainly be contrary to the predictions of models with $\Omega = 1$, especially models such as CHDM in which small-scale fluctuation power is significantly suppressed compared to standard CDM. For example. a possible detection of a large HI cloud was reported but not confirmed; the upgraded Arecibo telescope and the Giant Meter-wave Radio Telescope will soon provide sensitive tools for searching for such clouds at high redshift. Detections of galaxy clusters and quasar superclusters at redshifts $z \sim 2$ have also been reported, and there are some remarkable HST WFPC2 pictures of clusters at redshifts $z \sim 1$. Although the striking differences between the galaxies in such clusters and those at lower redshift support earlier indications that cluster galaxies have evolved significantly since redshift 1, galaxies in the field appear to have evolved less dramatically since the universe was half its present age (e.g., [24]). The most useful data would provide indications of the number densities of the objects (e.g. galaxies or clusters) considered at various red-shifts as a function of their mass (indicated for example by internal velocity dispersion or gas temperature), since that is what simulations predict most directly.

One of the most useful data sets for comparison with theories of structure formation is provided by absorption lines in the spectra of high-redshift quasars. The absorption systems with the highest density of neutral gas – known as damped Lyman $\alpha$ absorption systems – are presumably protogalaxies, and the quantity of gas in such systems at redshift $z \sim 3$ is roughly the same as the amount of ordinary matter in all the stars and gas that we can see in the universe at the present epoch (see e.g. [25]). An important question that will discriminate strongly between various cosmological models is whether the quantity of gas in such systems peaks at redshifts about 3-4, as expected in models where structure forms relatively late such as CHDM (see e.g. [26]), or increases to still higher redshift, as expected in models such as $\Lambda$CDM where structure forms significantly earlier.

## 5 Needs for the Future

Perhaps the most important information needed as a basis for constructing the first fundamental theory of cosmology is the values of the fundamental cosmological parameters, especially $H_0$, $\Omega$, and $\Lambda$. Below we summarize a number of other areas in which progress is needed.

*Bigger computers, faster and cleverer codes, shared software.* The greatest dynamical range in force resolution currently available is only a little better than three orders of magnitude in dissipationless simulations, worse in hydrodynamical simulations. This means that in a simulation with a 100 Mpc box, not large enough to simulate large surveys and large structures such as the Great Wall, the resolution is not much better than 100 kpc, an order of magnitude larger than the visible parts of galaxies. Moreover, there is a tradeoff between mass and force resolution: codes that permit better force resolution use fewer particles and thus have poorer mass resolution. A few groups such as the Grand Challenge Cosmology Consortium (GC$^3$) have recently devoted a great deal of effort to developing new and improved codes that exploit the new generation of massively parallel supercomputers that is now becoming available. It is very desirable that these codes become widely available so that the entire cosmology community can benefit. New technologies for visualization of the results of supercomputer calculations also hold considerable promise.

*Cosmic variance.* One of the worst problems with all simulations is cosmic variance; since only a random patch of universe is simulated, a number of such simulations must be compared to a number of regions for which galaxy survey data is available, and it is not clear that even the largest redshift surveys yet completed provide fair samples. This problem is exacerbated by the fact that, as many calculations have shown, there is a feed-down of effects from large structures to small; for example, rms pairwise velocities are strongly influenced by the presence of relatively rare rich clusters of galaxies (see e.g. [4]). Controlling cosmic variance is one of the most important challenges for current theory, until really large simulations can be compared to really large data sets such as the SDSS.

*Better understanding of processes leading to galaxy formation.* Since galaxy formation includes a number of generations of stars, including supernovae and the resulting chemical evolution, and perhaps often also involves more exotic objects such as the massive black holes thought to power active galactic nuclei, developing a secure understanding of these formation processes is likely to take a long time. It is even possible that a general theory of cosmology, including at least a general outline of the initial conditions and the nature of the dark matter, will precede rather than follow a detailed understanding of galaxy formation. No doubt a great deal of data on galaxies in both early and intermediate stages of formation will be necessary. Fortunately, such data is now coming from the new generation of great telescopes in space and on the ground. But present-epoch galaxies are brightest in the near infrared, and higher-redshift galaxies are expected to be brightest in the several micron band which can only be accessed from space. A large space infrared telescope such as SIRTF has long been seen as a high priority for astronomy, and the need for it was reiterated several times during the Snowmass workshop. Meanwhile, we will need to make use of even indirect data such as the amount of extra-galactic background infrared light from early galaxies, which can perhaps be probed by its absorption of TeV photons from AGNs via pair production [27].

*Better ways of comparing observational data with models.* We need new and better statistical tests, which are both *robust* against the difficulties of galaxy identification in simulations and the biases and selection effects always present in survey data, and *discriminatory* between the classes of cosmological models of interest. On the whole, it is probably better to compare theory with data by "observing" simulations rather than "correcting" data. It is very desirable that standard software become available to theorists and observers, so that standard versions of various statistics can be tried on many datasets from simulations and observations.

*Better access to observational and simulation data.* It is unfortunate that the only dense redshift survey covering a reasonably large volume which is publicly available is the 1982 CfA1 survey. Many papers have been published analyzing data from newer and larger redshift surveys in the years since then, but the redshift data remains largely unavailable. It is also desirable that simulation data (e.g. catalogs of objects identified as galaxies) be made available. The journals and funding agencies should ask a committee of observers and theorists to establish reasonable rules regarding access to such data – for example, all data used for a given paper must be made publicly accessible within one year of the publication of the paper – and ask referees to help enforce these rules. The POTENT group has set a good example of the sort of public access advocated here, by making their peculiar velocity dataset available in a timely way and in convenient form.